\documentclass[showpacs,showkeys,amssymb,aps,superscriptaddress,pre]{revtex4}
\usepackage{graphicx}
\usepackage{dcolumn}
\usepackage{bm}

\begin{document}


\title{Strong- and weak-universal critical behaviour of a mixed-spin Ising model with triplet interactions on the Union Jack (centered square) lattice}
\author{Jozef Stre\v{c}ka}
\affiliation{Department of Theoretical Physics and Astrophysics, Institute of Physics, Faculty of Science, \\ 
P. J. \v{S}af\'{a}rik University, Park Angelinum 9, 040 01 Ko\v{s}ice, Slovak Republic; jozef.strecka@upjs.sk}

\begin{abstract}
The mixed spin-1/2 and spin-$S$ Ising model on the Union Jack (centered square) lattice with four different three-spin (triplet) interactions and the uniaxial single-ion anisotropy is exactly solved by establishing a rigorous mapping equivalence with the corresponding zero-field (symmetric) eight-vertex model on a dual square lattice. A rigorous proof of the aforementioned exact mapping equivalence is provided by two independent approaches exploiting either a graph-theoretical or spin representation of the zero-field eight-vertex model. An influence of the interaction anisotropy as well as the uniaxial single-ion anisotropy on phase transitions and critical phenomena is examined in particular. It is shown that the considered model exhibits a strong-universal critical behaviour with constant critical exponents when considering the isotropic model with four equal triplet interactions or the anisotropic model with one triplet interaction differing from the other three. The anisotropic models with two different triplet interactions, which are pairwise equal to each other, contrarily exhibit a weak-universal critical behaviour with critical exponents continuously varying with a relative strength of the triplet interactions as well as the uniaxial single-ion anisotropy. It is evidenced that the variations of critical exponents of the mixed-spin Ising models with the integer-valued spins $S$ differ basically from their counterparts with the half-odd-integer spins $S$.
\end{abstract}

\maketitle

\section{Introduction}

One of the most important concepts elaborated in the theory of phase transitions and critical phenomena is universality hypothesis, which states that a critical behaviour does not depend on specific details of a model but only upon its spatial dimensionality, symmetry and number of components of the relevant order parameter. The foremost consequence of the universality hypothesis is that the critical behaviour of very different models may be characterized by the same set of critical exponents and one says that the models with the identical set of critical exponents belong to the same universality class. However, there exists a few exactly solved models whose critical exponents do depend on the interaction parameters and thus contradict the usual universality hypothesis \cite{bax82}. The spin-1/2 Ising model with a three-spin (triplet) interaction on planar lattices belongs to paradigmatic exactly solved models of this type. As a matter of fact, the exact solutions for the spin-1/2 Ising model with the triplet interaction gave rigorous proof for different sets of critical exponents  on different planar lattices \cite{hin72,bax73,bwu74,bax74,woo73,liu74}. More specifically, the critical exponent $\alpha$ for the specific heat fundamentally differs when this model is defined on centered square lattice ($\alpha = 1/2$) \cite{hin72}, triangular lattice ($\alpha = 2/3$) \cite{bax73,bwu74,bax74}, decorated triangular \cite{woo73}, honeycomb and diced lattices \cite{liu74} ($\alpha \approx 0$, logarithmic singularity). In addition, the spin-1/2 Ising model with the triplet interaction on a kagom\'e lattice \cite{bar89} does not display a phase transition at all.

In the present work we will consider and exactly solve the mixed spin-1/2 and spin-$S$ Ising model with the triplet interaction on the Union Jack (centered square) lattice by establishing a rigorous mapping correspondence with the symmetric (zero-field) eight-vertex model. The investigated model generalizes the model originally proposed and examined by Urumov \cite{uru80} when accounting for the additional uniaxial single-ion anisotropy acting on the spin-$S$ atoms. It will be demonstrated hereafter that the critical exponents of the mixed spin-1/2 and spin-$S$ Ising model with the triplet interaction on the centered square lattice fundamentally depend on the interaction anisotropy, the uniaxial single-ion anisotropy, as well as, the spin parity. 

\section{Model and exact solution}
\begin{figure}[h]
\centering
\includegraphics[width=5cm]{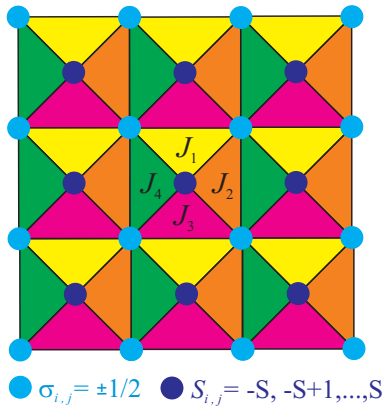}
\caption{A schematic illustration of the mixed spin-1/2 (light blue) and spin-$S$ (dark blue) Ising model on a centered square lattice. Four different colors are used in order to distinguish triplet interactions $J_1$, $J_2$, $J_3$ and $J_4$ within down-, left-, up- and right-pointing triangles, respectively.}
\label{csl}
\end{figure}

Let us introduce the mixed spin-1/2 and spin-$S$ Ising model with pure three-spin (triplet) interactions on a centered square lattice defined through the Hamiltonian:
\begin{eqnarray}
{\mathcal H} = - J_1 \!\sum_{i,j}^{\triangledown} \! S_{i,j} \sigma_{i,j} \sigma_{i+1,j} 
               - J_2 \!\sum_{i,j}^{\vartriangleleft} \! S_{i,j} \sigma_{i+1,j} \sigma_{i+1,j+1} 
               - J_3 \!\sum_{i,j}^{\vartriangle} \! S_{i,j} \sigma_{i,j+1} \sigma_{i+1,j+1} 
               - J_4 \!\sum_{i,j}^{\vartriangleright} \! S_{i,j} \sigma_{i,j} \sigma_{i,j+1} 
               - D \!\sum_{i,j} \! S_{i,j}^2, 
\label{ham} 
\end{eqnarray}
whereas the spin-1/2 atoms (light blue circles in Fig.~\ref{csl}) represented by the Ising spin variables $\sigma_{i,j}=\pm 1/2$ are placed at corners of a square lattice and the spin-$S$ atoms (dark blue circles in Fig.~\ref{csl}) are situated in the middle of square plaquettes. The Hamiltonian (\ref{ham}) takes into account four different triplet interactions $J_1$, $J_2$, $J_3$ and $J_4$ within down-, left-, up- and right-pointing triangles in addition to the uniaxial single-ion anisotropy $D$ acting on the spin-$S$ atoms. 
The Hamiltonian (\ref{ham}) can be alternatively rewritten as a sum of cell Hamiltonians ${\mathcal H} = \sum_{i,j}^{\square} {\mathcal H}_{i,j}$, whereas the cell Hamiltonian ${\mathcal H}_{i,j}$ involves all interactions terms depending on the central spin $S_{i,j}$: 
\begin{eqnarray} 
{\mathcal H}_{i,j}  = - J_1 S_{i,j} \sigma_{i,j} \sigma_{i+1,j} - J_2 S_{i,j} \sigma_{i+1,j} \sigma_{i+1,j+1} 
                      - J_3 S_{i,j} \sigma_{i,j+1} \sigma_{i+1,j+1} - J_4 S_{i,j} \sigma_{i,j} \sigma_{i,j+1} - D S_{i,j}^2. 
\end{eqnarray}
The partition function of the mixed spin-1/2 and spin-$S$ Ising model with triplet interactions on a centered square lattice can be then cast into the following form:
\begin{eqnarray}
{\cal{Z}} = \sum_{\{\sigma_{i,j}\}} \prod_{i,j} \sum_{S_{i,j}=-{\rm S}}^{\rm S} \exp(-\beta {\cal{H}}_{i,j}) 
          = \sum_{\{\sigma_{i,j}\}} \prod_{i,j} \omega (\sigma_{i,j},\sigma_{i+1,j}, \sigma_{i+1,j+1}, \sigma_{i,j+1}), 
\end{eqnarray}
where the summation $\sum_{\{\sigma_{i,j}\}}$ runs over all available spin configurations of the spin-1/2 atoms, $\beta = 1/(k_{\rm B} T)$, $k_{\rm B}$ is Boltzmann's constant, $T$ is the absolute temperature and the expression $\omega$ denotes the Boltzmann's weight obtained after tracing out degrees of freedom of the central spin-$S$ atom: 
\begin{eqnarray}
\omega (a, b, c, d) = \sum_{n=-{\rm S}}^{\rm S} \exp(\beta D n^2) \cosh \left[\beta n \left(J_1 a b + J_2 b c + J_3 c d + J_4 d a \right) \right].  
\label{bw}
\end{eqnarray}
\begin{figure}[h]
\includegraphics[width=12cm]{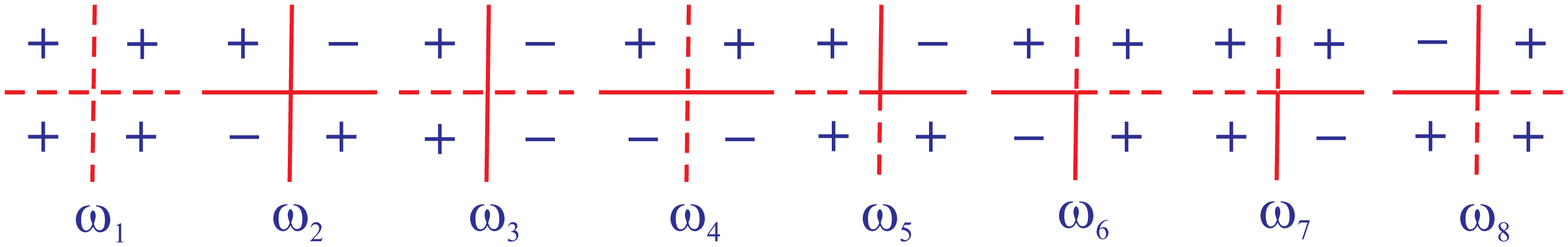}
\caption{A schematic representation of two-to-one mapping correspondence between the Ising spin configurations and line coverings of the equivalent eight-vertex model on a dual square lattice (the sign $\pm$ marks $\sigma = \pm 1/2$).}
\label{mc}
\end{figure}
An invariance of the Boltzmann's factor $\omega (a, b, c, d) = \omega (-a, -b, -c, -d)$ implies that there exist at most eight different Boltzmann's weights obtained from the expression (\ref{bw}) by considering all 16 spin configurations of the four corner spins involved therein. Hence, it follows that one may establish two-to-one mapping correspondence between a spin configuration and a relevant
graph representation of the equivalent eight-vertex model on a dual square lattice according to the scheme shown in Fig. \ref{mc}. A solid line is drawn on a respective edge of a dual square lattice lying in between two unequally aligned neighbouring spins, while a broken line is drawn otherwise. It turns out, moreover, that the effective Boltzmann's weights obtained after inserting all possible spin configurations of the four corner spins into the expression (\ref{bw}) are pairwise equal to each other: 
\begin{eqnarray} 
\omega_1 (+,+,+,+) \!\!\!\!\! &=& \!\!\!\!\! \omega_2 (+,-,+,-) 
= \sum_{n=-{\rm S}}^{\rm S} \exp(\beta D n^2) \cosh \left[\frac{\beta n}{4} \left(J_1 + J_2 + J_3 + J_4 \right) \right], \nonumber \\
\omega_3 (+,-,-,+) \!\!\!\!\! &=& \!\!\!\!\! \omega_4 (+,+,-,-) 
= \sum_{n=-{\rm S}}^{\rm S} \exp(\beta D n^2) \cosh \left[\frac{\beta n}{4} \left(J_1 - J_2 + J_3 - J_4 \right) \right], \nonumber \\
\omega_5 (-,+,+,+) \!\!\!\!\! &=& \!\!\!\!\! \omega_6 (+,+,-,+) 
= \sum_{n=-{\rm S}}^{\rm S} \exp(\beta D n^2) \cosh \left[\frac{\beta n}{4} \left(J_1 - J_2 - J_3 + J_4 \right) \right], \nonumber \\
\omega_7 (+,+,+,-) \!\!\!\!\!&=&\!\!\!\!\! \omega_8 (+,-,+,+) 
= \sum_{n=-{\rm S}}^{\rm S} \exp(\beta D n^2) \cosh \left[\frac{\beta n}{4} \left(J_1 + J_2 - J_3 - J_4 \right) \right],  
\label{ebw}
\end{eqnarray}
which means that the mixed-spin Ising model with triplet interactions on a centered square lattice is equivalent with the symmetric (zero-field) eight-vertex model exactly solved by Baxter \cite{bax71,bax72}. Owing to this fact, one may easily prove an exact mapping relationship between the partition functions of the mixed-spin Ising model with triplet interactions on a centered square lattice and the zero-field eight-vertex model on a dual square lattice:
\begin{eqnarray}
{\cal{Z}} (\beta, J_1, J_2, J_3, J_4, D) = 2 {\cal Z}_{8-{\rm vertex}} (\omega_1, \omega_3, \omega_5, \omega_7). 
\label{pf}           
\end{eqnarray}
It is apparent from the mapping relation (\ref{pf}) between the partition functions that the mixed-spin Ising model with triplet interactions on a centered square lattice becomes critical only if the corresponding zero-field eight-vertex model becomes critical as well. Bearing this in mind, the critical points of the mixed-spin Ising model with triplet interactions on a centered square lattice can be readily obtained from Baxter's critical condition \cite{bax71,bax72} when the explicit form for the effective Boltzmann's weights (\ref{ebw}) is taken into consideration:
\begin{eqnarray}
\omega_1+\omega_3+\omega_5+\omega_7=2{\rm max} \{\omega_1,\omega_3,\omega_5,\omega_7\}.
\label{cc}           
\end{eqnarray}
It should be stressed that the critical exponents for the specific heat, magnetization, susceptibility and correlation length satisfy Suzuki’s weak-universal hypothesis \cite{suz74} and can be calculated from: 
\begin{eqnarray}
\alpha = \alpha' = 2 - \pi/\mu, \, \, \, \quad \beta = \pi/16 \mu, \, \, \, \quad \gamma = \gamma' = 7 \pi/8 \mu, \, \, \, \quad 
\nu = \nu' = \pi/2 \mu, 
\label{ce}
\end{eqnarray}
where $\tan (\mu/2) = (\omega_1 \omega_3/ \omega_5 \omega_7)^{1/2}$ on assumption that $\omega_1 = {\rm max} \{\omega_1,\omega_3,\omega_5,\omega_7\}$.

\begin{figure}
\includegraphics[width=8cm]{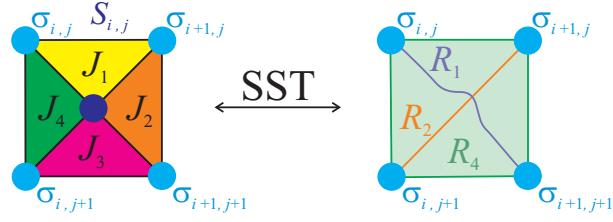}
\caption{A schematic representation of the generalized star-square transformation, which replaces spin degrees of freedom of the central spin-$S$ atom through two effective pair interactions ($R_1$, $R_2$) and the effective quartic interaction ($R_4$) between the four enclosing spin-1/2 atoms.}
\label{sstf}
\end{figure}  

The exact mapping equivalence with the zero-field eight-vertex model can be alternatively proven by exploiting the spin representation of the eight-vertex model. For this purpose, the effective Boltzmann factor (\ref{bw}) can be replaced via the generalized star-square transformation \cite{fis59,roj09,str10} schematically drawn in Fig. \ref{sstf}:
\begin{eqnarray}
\omega \!\!\!&=&\!\!\! \sum_{n=-{\rm S}}^{\rm S} \!\!\! \exp(\beta D n^2) \cosh \Bigl[\beta n (J_1 \sigma_{i,j} \sigma_{i+1,j} + J_2 \sigma_{i+1,j} \sigma_{i+1,j+1}                + J_3 \sigma_{i+1,j+1} \sigma_{i,j+1} + J_4 \sigma_{i,j+1} \sigma_{i,j}) \Bigr] \nonumber \\
\!\!\!&=&\!\!\! R_0 \exp(\beta R_1 \sigma_{i,j} \sigma_{i+1,j+1} + \beta R_2 \sigma_{i+1,j} \sigma_{i,j+1} 
+ \beta R_4 \sigma_{i,j} \sigma_{i+1,j} \sigma_{i+1,j+1} \sigma_{i,j+1}). 
\label{sst}
\end{eqnarray}
The physical meaning of the generalized star-square transformation (\ref{sst}) lies in replacing spin degrees of freedom related to the central spin-$S$ atom through the effective pair interactions ($R_1$, $R_2$) and the effective quartic interaction ($R_4$) between the four enclosing spin-1/2 atoms (see Fig. \ref{sstf}). The star-square transformation (\ref{sst}) must hold for any spin state of the four enclosing spin-1/2 atoms and this 'self-consistency' condition unambiguously determines so far unspecified mapping parameters:
\begin{eqnarray}
R_0 = \left(\omega_1 \omega_3 \omega_5 \omega_7 \right)^{1/4}\!, \,\,\,
\beta R_1 = \ln \left( \frac{\omega_1 \omega_7}{\omega_3 \omega_5} \right)\!, \,\,\,
\beta R_2 = \ln \left( \frac{\omega_1 \omega_5}{\omega_3 \omega_7} \right)\!,  \,\,\,
\beta R_4 = 4 \ln \left( \frac{\omega_1 \omega_3}{\omega_5 \omega_7} \right). 
\end{eqnarray} 
The star-square transformation (\ref{sst}) establishes a rigorous mapping correspondence between the partition function of the mixed-spin Ising model with triplet interactions on a centered square lattice and the partition function of the spin-1/2 Ising model on two inter-penetrating square lattices with the effective pair interactions ($R_1$, $R_2$) and the effective quartic interaction ($R_4$):
\begin{eqnarray}
{\cal Z} (\beta, J_1, J_2, J_3, J_4, D) = R_0^{2N} {\cal Z}_{8-{\rm vertex}} (\beta, R_1, R_2, R_4).
\nonumber
\end{eqnarray} 
It has been proved previously that the spin-1/2 Ising model defined on two inter-penetrating square lattices coupled together by means of the quartic interaction is nothing but the Ising representation of the zero-field eight-vertex model on a square lattice \cite{wu71,kad71}. In this way we have afforded alternative proof for an exact mapping equivalence between the mixed-spin Ising model with triplet interactions on a centered square lattice and the zero-field eight-vertex model on a square lattice.

\section{Results and discussion}

\begin{figure}[h]
\centering
\includegraphics[width=3cm]{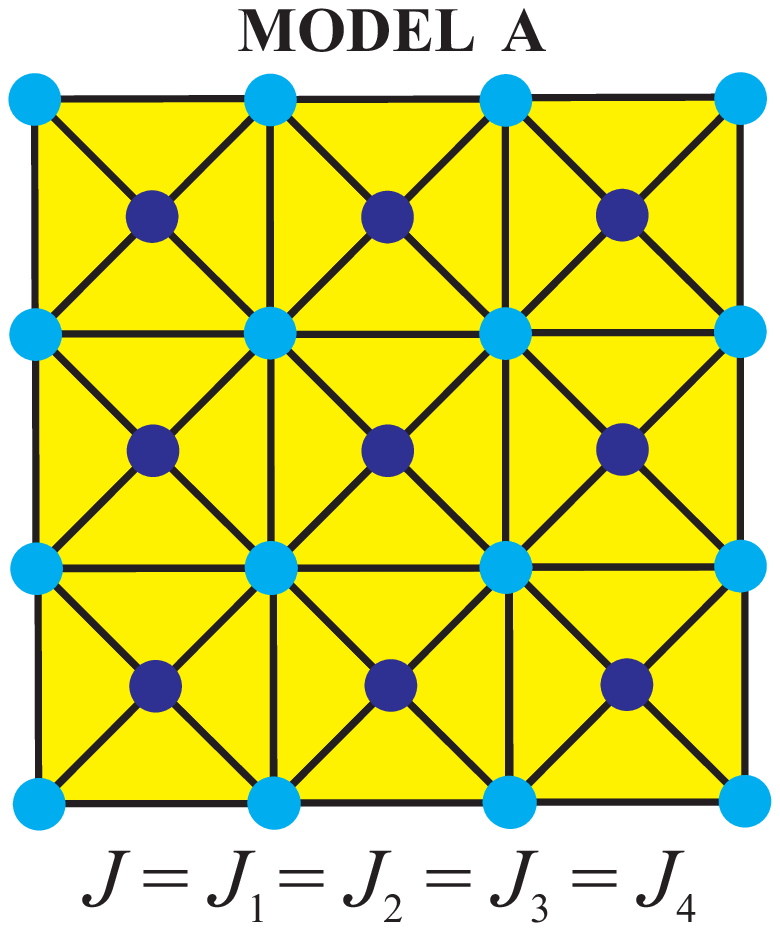}
\hspace{0.5cm}
\includegraphics[width=3cm]{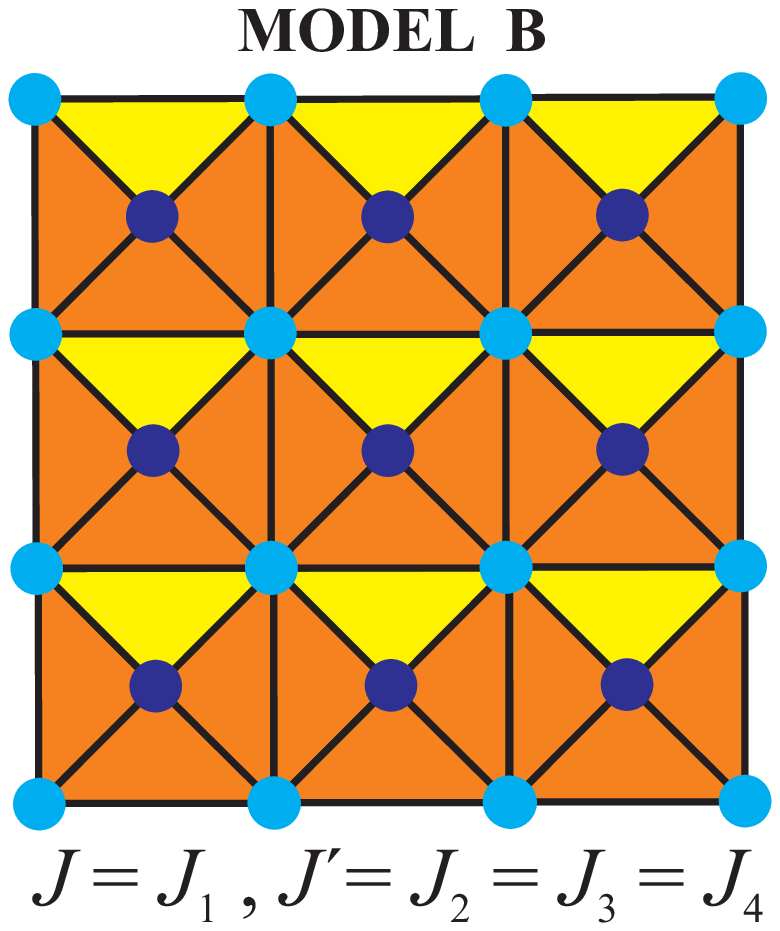}
\hspace{0.5cm}
\includegraphics[width=3cm]{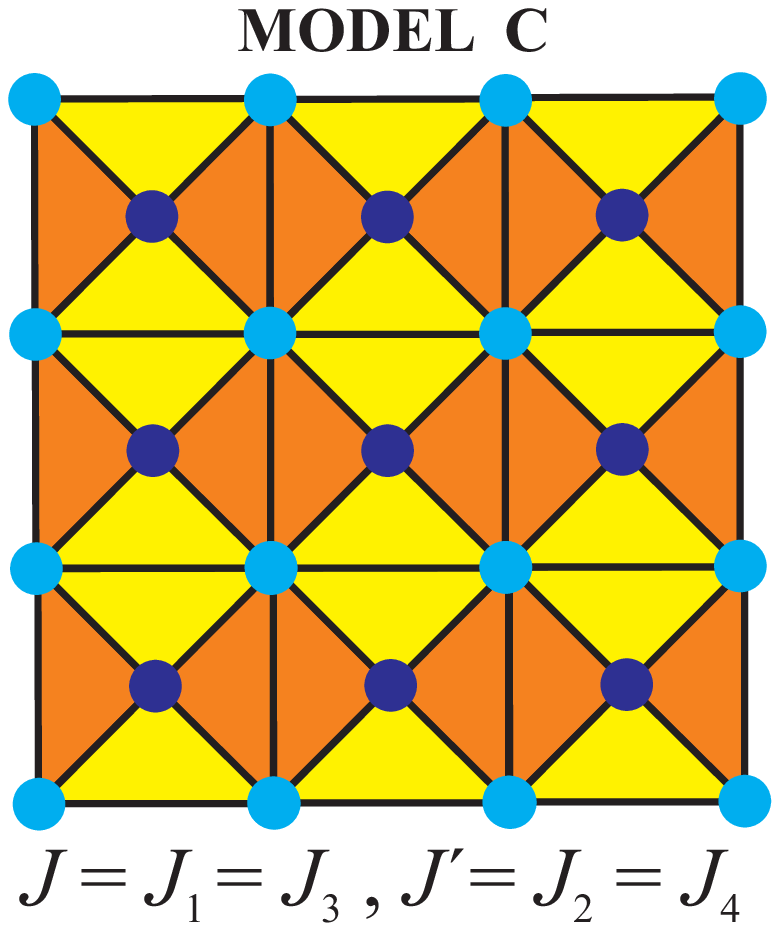}
\hspace{0.5cm}
\includegraphics[width=3cm]{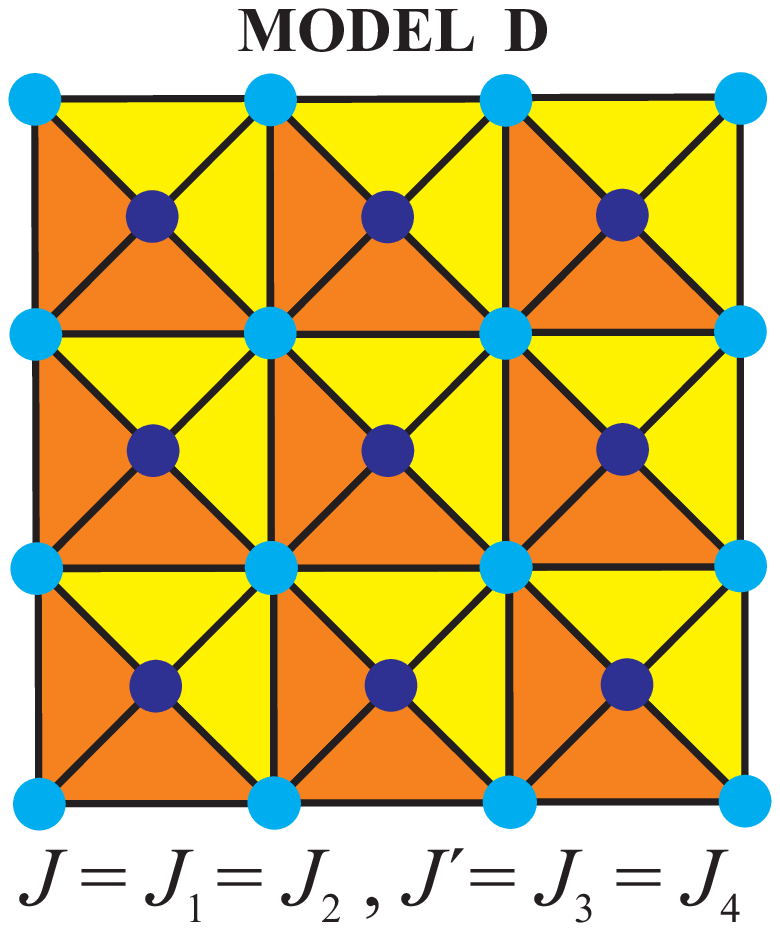}
\caption{The four particular cases of the mixed spin-$1/2$ and spin-$S$ Ising model with triplet interactions on a centered square lattice, which will be comprehensively studied in the following subsections. Different colors are used for distinguishing triplet interactions of different size.}
\label{models}
\end{figure}

In this section, let us discuss the most interesting results for the mixed spin-$1/2$ and spin-$S$ Ising model with triplet interactions on a centered square lattice depending on the interaction anisotropy, the uniaxial single-ion anisotropy and the spin magnitude $S$. For the sake of simplicity, our further attention will be restricted to four particular cases to be further referred to as:
\begin{itemize}
\item Model A: $J \equiv J_1 = J_2 = J_3 = J_4$,
\item Model B: $J \equiv J_1, J' \equiv J_2 = J_3 = J_4$,
\item Model C: $J \equiv J_1 = J_3, J' \equiv J_2 = J_4$,
\item Model D: $J \equiv J_1 = J_2, J' \equiv J_3 = J_4$,
\end{itemize}
which will be separately treated in the following subsections. For better illustration, the four aforementioned special cases of the mixed spin-$1/2$ and spin-$S$ Ising model with triplet interactions on a centered square lattice are schematically drawn in Fig. \ref{models}, where different colors are used for distinguishing triplet interactions of different size. 

\subsection{Model A $(J \equiv J_1 = J_2 = J_3 = J_4)$}
\begin{figure}[h]
\centering
\includegraphics[width=14cm]{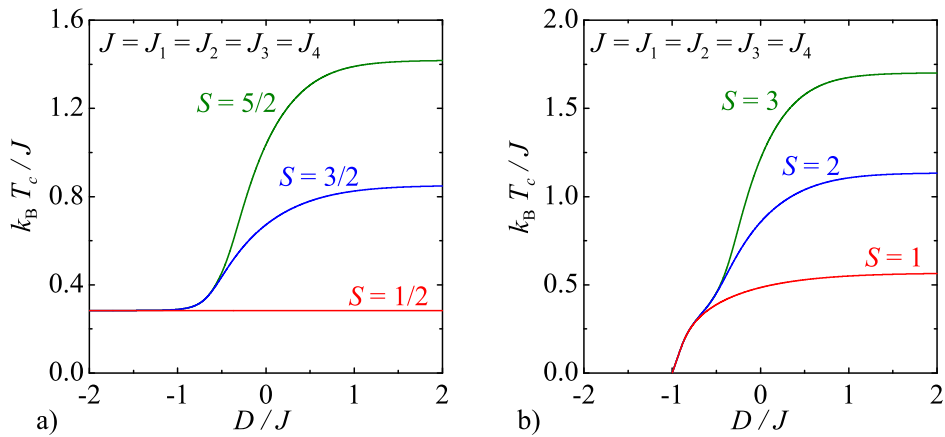}
\vspace{-1.0cm}
\caption{The critical temperature of the model A as a function of the uniaxial single-ion anisotropy by considering: (a) half-odd-integer spins $S$; (b) integer spins $S$.}
\label{modela}
\end{figure}

Let us begin with a detailed analysis of critical behaviour of the mixed spin-$1/2$ and spin-$S$ Ising model with unique triplet interaction on a centered square lattice, which represents a very special case due to the isotropic nature of the triplet interactions $J \equiv J_1 = J_2 = J_3 = J_4$. Under this condition, one gains from Eq. (\ref{ebw}) just two different Boltzmann's weights:
\begin{eqnarray} 
\omega_1 = \sum_{n=-{\rm S}}^{\rm S} \exp(\beta D n^2)  \cosh \left(\beta n J \right), \qquad
\omega_3 = \omega_5 = \omega_7 = \sum_{n=-{\rm S}}^{\rm S} \exp(\beta D n^2),  
\label{bwa}
\end{eqnarray}
whereas the Boltzmann's weights (\ref{bwa}) evidently satisfy the inequality $\omega_1 \geq \omega_3$. In this regard, the critical condition (\ref{cc}) of the mixed spin-$1/2$ and spin-$S$ Ising model with unique triplet interaction on a centered square lattice simplifies to:
\begin{eqnarray} 
\omega_1 = 3 \omega_3 \quad \Rightarrow \quad
\sum_{n=-{\rm S}}^{\rm S} \exp(\beta_{\rm c} D n^2) 
\cosh \left(\beta_{\rm c} n J \right) = 3 \sum_{n=-{\rm S}}^{\rm S} \exp(\beta_{\rm c} D n^2), 
\label{cca}
\end{eqnarray}
where $\beta_{\rm c} = 1/(k_{\rm B} T_{\rm c})$ and $T_{\rm c}$ marks the critical temperature. It follows from Eq. (\ref{ce}) that the critical exponents remain constant along the whole critical line (\ref{cca}) irrespective of the spin magnitude $S$: 
\begin{eqnarray} 
\tan(\mu/2) = \sqrt{\omega_1/\omega_3} = \sqrt{3} \quad \Rightarrow \quad \alpha = \alpha' = 1/2, \, \, \, \, \beta = 3/32, \, \, \, \, \gamma = \gamma' = 21/16, \, \, \, \, \nu =\nu' = 3/4, 
\label{cea}
\end{eqnarray}
whereas their size is identical with the ones predicted for the spin-1/2 Ising model with the unique triplet interaction on a centered square lattice, i.e. the so-called Hintermann-Merlini model \cite{hin72}. The critical temperature obtained from the numerical solution of the critical condition (\ref{cca}) is plotted in Fig. \ref{modela} against the uniaxial single-ion anisotropy for several spin magnitudes $S$. Although the critical temperature monotonically decreases with decreasing of the uniaxial single-ion anisotropy regardless of the spin size $S$, there is fundamental difference in the critical behaviour of the mixed-spin Ising models with integer and half-odd-integer spins $S$, respectively. Namely, the critical temperature of the former mixed-spin systems becomes zero for $D/J < -1$ in accordance with presence of the disordered ground state, which appears due to energetic favouring of the nonmagnetic spin state $S=0$ of the integer-valued spins. On the other hand, the critical temperature of the latter mixed-spin systems tends towards the critical temperature of the Hintermann-Merlini model \cite{hin72} $k_{\rm B} T_{\rm c}/J = 1/4\ln(1+\sqrt{2}) \approx 0.2836\ldots$, which is achieved in the asymptotic limit $D/J \to -\infty$ (but practically already at $D/J \approx -1$) due to energetic favouring of two lowest-valued states $S=\pm 1/2$ of the half-odd-integer spins. 
                       
\subsection{Model B $(J \equiv J_1, J' \equiv J_2 = J_3 = J_4)$}
Next, let us relax the condition of the isotropic triplet interactions by considering the model B, where one triplet interaction (say $J\equiv J_1$) differs from the other three ($J' \equiv J_2 = J_3 = J_4$). Even under this constraint one still gets from Eq. (\ref{ebw}) just two different Boltzmann's weights: 
\begin{eqnarray} 
\omega_1 = \!\!\! \sum_{n=-{\rm S}}^{\rm S} \!\!\! \exp(\beta D n^2)  \cosh \left[ \frac{\beta n}{4} \left(J + 3J'\right) \right]\!, \, \, \, 
\omega_3 = \omega_5 = \omega_7 = \!\!\! \sum_{n=-{\rm S}}^{\rm S} \!\!\! \exp(\beta D n^2)  \cosh \left[ \frac{\beta n}{4} \left(J - J'\right) \right]\!\!,  
\label{ccb}
\end{eqnarray}
which are however slightly more complicated due to the interaction anisotropy. It is evident from Eq. (\ref{ccb}) that the Boltzmann's weights still satisfy the inequality $\omega_1 \geq \omega_3$, which affords the following critical condition: 
\begin{eqnarray} 
\omega_1 = 3 \omega_3 \quad \Rightarrow 
\sum_{n=-{\rm S}}^{\rm S} \!\!\! \exp(\beta_{\rm c} D n^2) \cosh \left[ \frac{\beta_{\rm c} n}{4} \left(J + 3J'\right) \right]
 = 3 \!\!\! \sum_{n=-{\rm S}}^{\rm S} \!\!\! \exp(\beta_{\rm c} D n^2)  \cosh \left[ \frac{\beta_{\rm c} n}{4} \left(J - J'\right) \right]\!\!. 
\end{eqnarray}
It should be pointed out, moreover, that the critical exponents are still constants independent of the spin magnitude and the interaction anisotropy as given by Eq. (\ref{cea}). 

\begin{figure}[h]
\centering
\includegraphics[width=14cm]{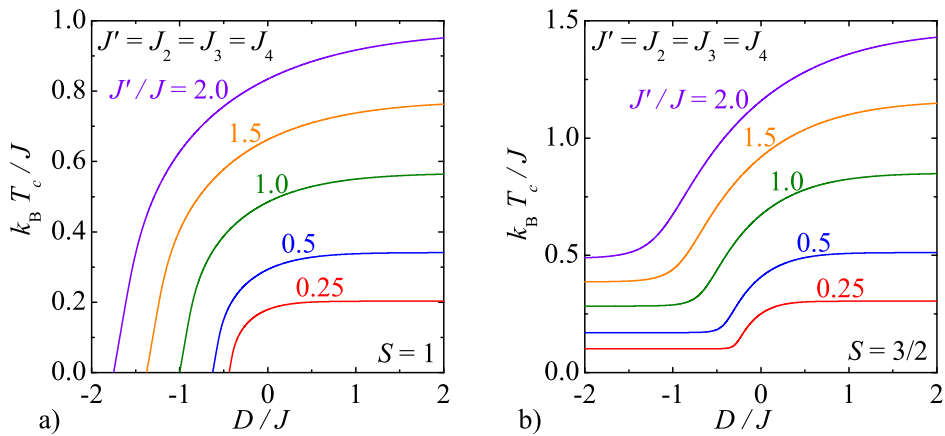}
\vspace{-1.0cm}
\caption{The critical temperature of the model B as a function of the uniaxial single-ion anisotropy by considering several values of the  interaction anisotropy $J'/J$ and two different spin magnitudes: (a) $S=1$; (b) $S=3/2$.}
\label{modelb}
\end{figure}

The critical frontiers of the model B are illustrated in Fig. \ref{modelb} for two different spin values, which demonstrate typical critical behaviour of the mixed-spin systems with half-odd-integer and integer spins $S$, respectively. It can be seen from Fig. \ref{modelb} that the critical temperature of the investigated mixed-spin system rises steadily with increasing of the interaction ratio $J'/J$ both for the integer as well as half-odd-integer spins. However, it is worthwhile to remark that the critical value of the uniaxial single-ion anisotropy needed for an onset of the disordered ground state of the mixed-spin systems with integer spins $S$ shifts towards more negative values upon strengthening of the interaction ratio $J'/J$.    

\subsection{Model C ($J \equiv J_1 = J_3, J' \equiv J_2 = J_4$)}
Now, let us turn our attention to a critical behaviour of the model C with the triplet interactions $J \equiv J_1 = J_3$, $J' \equiv J_2 = J_4$, which are pairwise equal in triangles lying in opposite to each other within elementary square cells (see Fig. \ref{models}). In this particular case one gets from Eq. (\ref{ebw}) three different Boltzmann's weights: 
\begin{eqnarray} 
\omega_1 \!\!\!&=&\!\!\! \!\!\! \sum_{n=-{\rm S}}^{\rm S} \!\!\! \exp(\beta D n^2)  \cosh \left[ \frac{\beta n}{2} \left(J + J'\right) \right]\!, \qquad 
\omega_3 = \sum_{n=-{\rm S}}^{\rm S} \!\!\! \exp(\beta D n^2)  \cosh \left[ \frac{\beta n}{2} \left(J - J'\right) \right]\!,  \nonumber \\ 
\omega_5 \!\!\!&=&\!\!\! \omega_7 = \!\!\! \sum_{n=-{\rm S}}^{\rm S} \!\!\! \exp(\beta D n^2).  
\label{bwc}
\end{eqnarray}
It directly follows from Eq. (\ref{bwc}) that the Boltzmann's weights obey the inequality $\omega_1 \geq \omega_3 \geq \omega_5$ and hence, the critical condition $\omega_1 = \omega_3 + 2 \omega_5$ can be explicitly written as follows: 
\begin{eqnarray} 
\!\!\! \sum_{n=-{\rm S}}^{\rm S} \!\!\! \exp(\beta_{\rm c} D n^2) \cosh \! \left[ \frac{\beta_{\rm c} n}{2} \! \left(J + J'\right) \right] 
= \!\!\! \sum_{n=-{\rm S}}^{\rm S} \!\!\! \exp(\beta_{\rm c} D n^2) \cosh \! \left[ \frac{\beta_{\rm c} n}{2} \! \left(J - J'\right) \right] 
+ 2 \!\!\! \sum_{n=-{\rm S}}^{\rm S} \!\!\! \exp(\beta_{\rm c} D n^2). 
\label{ccc}
\end{eqnarray}
Besides, the Boltzmann's weights (\ref{bwc}) imply that the critical exponents (\ref{ce}) along the critical line (\ref{ccc}) may display striking dependence on the spin magnitude, the uniaxial single-ion anisotropy as well as the interaction anisotropy according to the formulas: 
\begin{eqnarray} 
\tan \left( \frac{\mu}{2} \right)=\frac{\sqrt{\omega_1 \omega_3}}{\omega_5} \quad \Rightarrow \quad 
\alpha = \alpha' = 2 - \frac{\pi}{\mu}, \quad \beta = \frac{\pi}{16 \mu}, \quad \gamma = \gamma' = \frac{7 \pi}{8 \mu}, \quad
\nu = \nu' = \frac{\pi}{2 \mu}. 
\label{cec}
\end{eqnarray}

\begin{figure}[h]
\centering
\includegraphics[width=14cm]{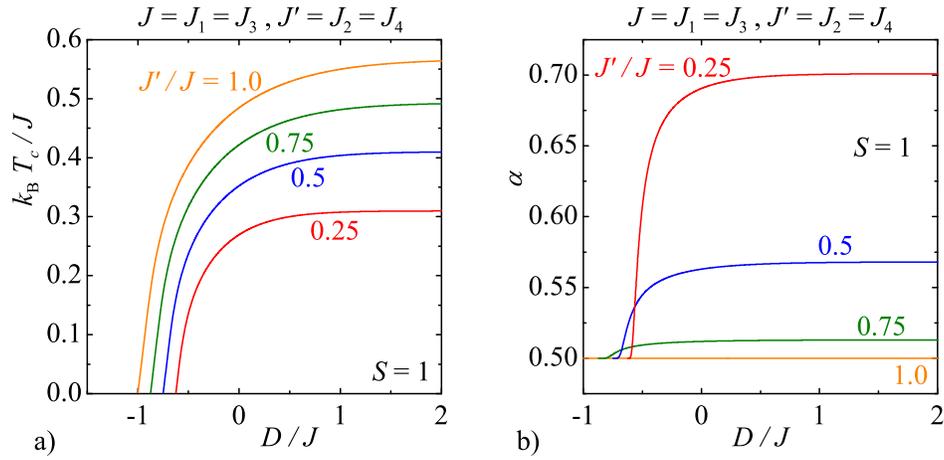}
\vspace{-1.0cm}
\caption{(a) The critical temperature of the model C as a function of the uniaxial single-ion anisotropy by considering the spin size $S=1$ and several values of the interaction anisotropy $J'/J$; (b) The respective changes of the critical exponent $\alpha$ for the specific heat along the critical lines displayed in Fig. \ref{modelc1}(a).}
\label{modelc1}
\end{figure}

\begin{figure}[h]
\centering
\includegraphics[width=14cm]{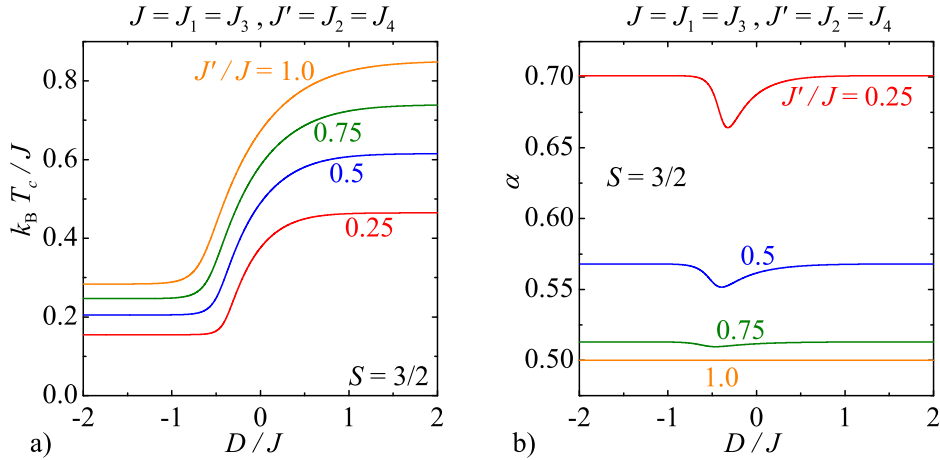}
\vspace{-1.0cm}
\caption{(a) The critical temperature of the model C as a function of the uniaxial single-ion anisotropy by considering the spin size $S=3/2$ and several values of the interaction anisotropy $J'/J$; (b) The respective changes of the critical exponent $\alpha$ for the specific heat along the critical lines displayed in Fig. \ref{modelc32}(a).}
\label{modelc32}
\end{figure}	

For illustrative purposes, Fig. \ref{modelc1}(a) and Fig. \ref{modelc32}(a) display phase boundaries of the model C for the specific spin values $S=1$ and $3/2$, respectively. As one can see, the same general trends can be observed in the relevant dependencies of the critical temperature on the uniaxial single-ion anisotropy and the interaction anisotropy as previously discussed for the model B. However, the model C exhibits along the displayed critical lines continuously varying critical exponents unlike the model B with the strong-universal (constant) critical exponents. For instance, it can be found from Fig. \ref{modelc1}(b) that the critical exponent $\alpha$ for the specific heat monotonically increases with increasing of the uniaxial single-ion anisotropy for the integer-valued spins $S=1$, whereas the observed increase is the greater the higher the interaction anisotropy is. In addition, it is quite surprising that the critical exponents for the model C with the half-odd-integer spins $S=3/2$ exhibit completely different weak-universal critical behaviour. Although the critical exponent $\alpha$ for the specific heat is shifted towards higher values upon increasing of the interaction anisotropy too, but this time the critical exponent $\alpha$ displays a more peculiar nonmonotonous dependence on the uniaxial single-ion anisotropy with a round minimum whose depth basically depends on the interaction anisotropy.

\subsection{Model D $(J \equiv J_1 = J_2, J' \equiv J_3 = J_4)$}

Last but not least, we will perform a comprehensive analysis of critical behaviour of the model D with the triplet interactions $J \equiv J_1 = J_2$, $J' \equiv J_3 = J_4$, which are pairwise equal to each other within adjacent triangles of elementary square plaquettes. In this particular case one gets from Eq. (\ref{ebw}) three different Boltzmann's weights given by:
\begin{eqnarray} 
\omega_1 \!\!\!&=&\!\!\! \!\!\! \sum_{n=-{\rm S}}^{\rm S} \!\!\! \exp(\beta D n^2)  \cosh \left[ \frac{\beta n}{2} \left(J + J'\right) \right]\!,  \qquad
\omega_3 = \omega_5 = \sum_{n=-{\rm S}}^{\rm S} \!\!\! \exp(\beta D n^2), \nonumber  \\ 
\omega_7 \!\!\!&=&\!\!\! \!\!\! \sum_{n=-{\rm S}}^{\rm S} \!\!\! \exp(\beta D n^2)  \cosh \left[ \frac{\beta n}{2} \left(J - J'\right) \right]\!.  
\label{bwd} 
\end{eqnarray}
It is quite apparent that the Boltzmann's weights (\ref{bwd}) satisfy the inequality $\omega_1 \geq \omega_7 \geq \omega_3$, which affords from Eq. (\ref{cc}) the critical condition $\omega_1 = 2 \omega_3 + \omega_7$. After substituting explicit form of the Boltzmann's weights (\ref{bwd}) into the relevant critical condition one obtains the completely identical critical condition as given by Eq. (\ref{ccc}) for the model C. This result would suggest that the model D has exactly the same phase boundaries as the model C. However, it should be emphasized that the weak-universal critical behaviour of the model D will be characterized by different critical exponents: 
\begin{eqnarray} 
\tan \left( \frac{\mu}{2} \right) = \sqrt{\frac{\omega_1}{\omega_7}} 
\quad \Rightarrow \quad  
\alpha = \alpha' = 2 - \frac{\pi}{\mu}, \quad \beta = \frac{\pi}{16 \mu}, \quad \gamma = \gamma' = \frac{7 \pi}{8 \mu} \quad
\nu = \nu' = \frac{\pi}{2 \mu},
\label{ced} 
\end{eqnarray}
because the definition of the parameter $\mu$ governing changes of the critical exponents is different [c.f. Eqs. (\ref{cec}) and (\ref{ced})]. 

\begin{figure}[h]
\centering
\includegraphics[width=14cm]{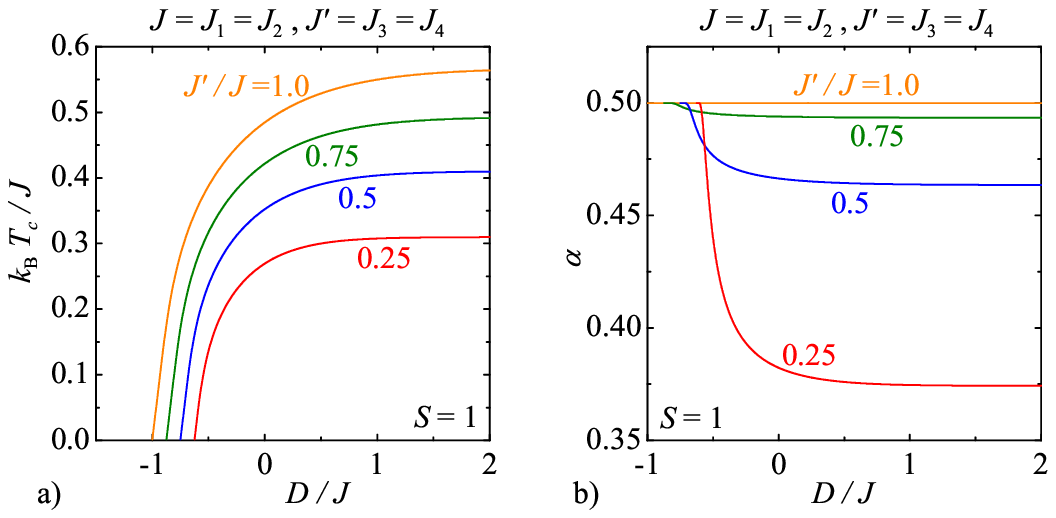}
\vspace{-1.0cm}
\caption{(a) The critical temperature of the model D as a function of the uniaxial single-ion anisotropy by considering the spin size $S=1$ and several values of the interaction anisotropy $J'/J$; (b) The respective changes of the critical exponent $\alpha$ for the specific heat along the critical lines displayed in Fig. \ref{modeld1}(a).}
\label{modeld1}
\end{figure}
\begin{figure}[h]
\centering
\includegraphics[width=14cm]{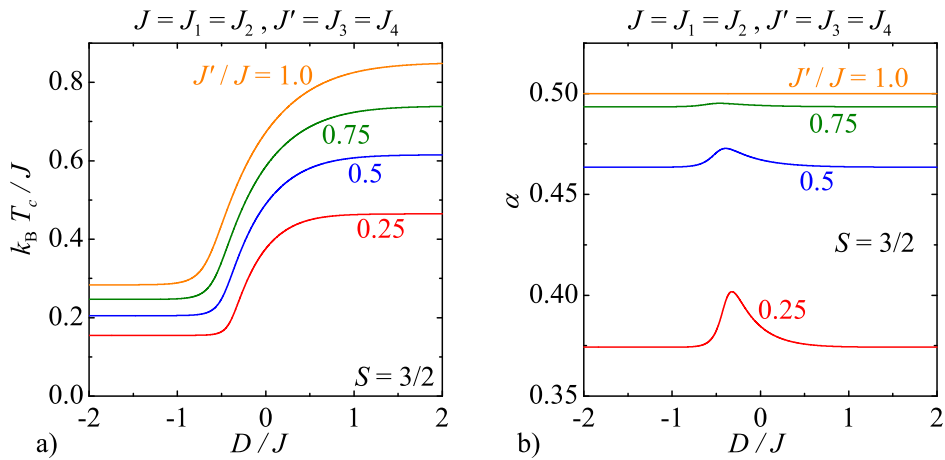}
\vspace{-1.0cm}
\caption{(a) The critical temperature of the model D as a function of the uniaxial single-ion anisotropy by considering the spin size $S=3/2$ and several values of the interaction anisotropy $J'/J$; (b) The respective changes of the critical exponent $\alpha$ for the specific heat along the critical lines displayed in Fig. \ref{modeld32}(a).}
\label{modeld32}
\end{figure}

For the sake of comparison, the critical temperature and critical exponent $\alpha$ are plotted in Figs. \ref{modeld1} and \ref{modeld32} for the model D by assuming two different spin values $S=1$ and $3/2$, respectively. It has already been argued that the phase boundaries of the model D coincide with the critical lines of the model C, so let us only comment the respective behaviour of the critical exponent $\alpha$ for the specific heat. Fig. \ref{modeld1}(b) would suggest that the critical exponent $\alpha$ for the model D with the integer-valued spin $S=1$ falls down monotonically with increasing of the uniaxial single-ion anisotropy, whereas the interaction anisotropy generally reinforces the relevant decline. This behaviour is in sharp contrast with what has been previously reported for the model C, where exactly opposite tendency has been revealed [c.f. Fig. \ref{modeld1}(b) with Fig. \ref{modelc1}(b)]. The reduction of the critical exponent $\alpha$ due to the interaction anisotropy has also been detected for the model D with the half-odd-integer spin $S=3/2$, but this time the critical exponent $\alpha$ shows a peculiar nonmonotonous dependence on the uniaxial single-ion anisotropy with a round maximum. This finding is repeatedly in marked contrast with what has been previously found for the model C [c.f. Fig. \ref{modeld32}(b) with Fig. \ref{modelc32}(b)]. It could be thus concluded that the models C and D display remarkably different weak-universal critical behaviour of the critical exponents in spite of the fact that their phase boundaries are in a perfect coincidence.

\section{Conclusions}

The mixed spin-1/2 and spin-$S$ Ising model with four different triplet interactions on a centered square lattice has been exactly solved by establishing a rigorous mapping equivalence with the corresponding zero-field  eight-vertex model on a dual square lattice. A rigorous proof of the aforementioned exact mapping equivalence has been afforded by two independent ways exploiting either the graph-theoretical formulation \cite{bax71,bax72} or the spin representation \cite{wu71,kad71} of the zero-field eight-vertex model, the latter one obtained after adapting of the generalized star-square mapping transformation \cite{fis59,roj09,str10}.  

The primary goal of the present work was to examine an influence of the interaction anisotropy, the uniaxial single-ion anisotropy and the spin parity upon phase transitions and critical phenomena. It has been shown that the considered model exhibits a strong-universal critical behaviour with constant critical exponents, which are independent of the uniaxial single-ion anisotropy as well as the spin parity when considering either the isotropic model A with four equal triplet interactions or the anisotropic model B with one triplet interaction differing from the other three. On the other hand, it has also been evidenced that the models C and D with the triplet interactions, which are pairwise equal to each other, exhibit a weak-universal critical behaviour characterized by continuously varying critical exponents. Under these circumstances, the relevant critical exponents are changing along the critical lines in dependence on a relative strength of the triplet interactions as well as the uniaxial single-ion anisotropy. Although the critical boundaries of the models C and D are completely the same, different interaction anisotropy of both models is responsible for a qualitatively different weak-universal behaviour of the critical exponents. Besides, it also turns out that the mixed-spin systems with the integer-valued spins $S$ exhibit very different variations of the critical exponents in comparison with their counterparts with the half-odd-integer spins $S$. 

\begin{acknowledgments}
This work was financially supported by the grant of The Ministry of Education, Science, Research and Sport of the Slovak Republic under the contract No. VEGA 1/0331/15 and by the grant of the Slovak Research and Development Agency under the contract No. APVV-16-0186.
\end{acknowledgments}




\begin{thebibliography}{100}
\bibitem{bax82} Baxter R.J., {\em Exactly Solved Models in Statistical Mechanics}; Academic: New York, 1982.  
\bibitem{hin72} Hintermann, A.; Merlini, D. Exact solution of a two dimensional Ising model with pure 3 spin interactions. 
\textit{Phys. Lett.} \textbf{1972}, \textit{41A}, 208-210. 
\bibitem{bax73} Baxter, R.J.; Wu, F.Y. Exact Solution of an Ising Model with Three-Spin Interactions on a Triangular Lattice.
\textit{Phys. Rev. Lett.} \textbf{1973}, \textit{31}, 1294-1297.
\bibitem{bwu74} Baxter, R.J.; Wu, F.Y. Ising Model on a Triangular Lattice with Three-spin Interactions. I. The Eigenvalue Equation.
\textit{Aust. J. Phys.} \textbf{1974}, \textit{27}, 357-368. 
\bibitem{bax74} Baxter, R.J. Ising Model on a Triangular Lattice with Three-spin Interactions. II. Free Energy and Correlation Length. 
\textit{Aust. J. Phys.} \textbf{1974}, \textit{27}, 369-382. 
\bibitem{woo73} Wood, D.W. A free-fermion model, and the solution of an Ising model with pure triplet interactions. 
\textit{J. Phys. C: Solid St. Phys.} \textbf{1973}, \textit{6}, L135-L138. 
\bibitem{liu74} Liu, L.L.; Stanley, H.E. Exact solution in an external magnetic field of Ising models with three-spin interactions. 
\textit{Phys. Rev. B} \textbf{1974}, \textit{10}, 2958-2961.
\bibitem{bar89} Barry, J.H.; Wu, F.Y. Exact Solutions For a 4-Spin-interaction Ising Model on the d=3 Pyrochlore Lattice. 
\textit{Int. J. Mod. Phys.} \textbf{1989}, \textit{3}, 1247.
\bibitem{uru80} Urumov, V. Examples of Spin and External Field Dependent Critical Exponents. 
In {\em Ordering in two dimensions}; Sinha S.K., Ed.; Elsevier: New York, USA, 1980; pp. 361-364.
\bibitem{bax71} Baxter, R.J. Eight-Vertex Model in Lattice Statistics. \textit{Phys. Rev. Lett.} \textbf{1971}, \textit{26}, 832-833. 
\bibitem{bax72} Baxter, R.J. Partition function of the Eight-Vertex lattice model. \textit{Ann. Phys.} \textbf{1972}, \textit{70}, 193-228.
\bibitem{suz74} Suzuki, M. New Universality of Critical Exponents. \textit{Prog. Theor. Phys.} \textbf{1974}, \textit{51}, 1992-1993.
\bibitem{fis59} Fisher, M.E. Transformations of Ising Models. \textit{Phys. Rev.} \textbf{1959}, \textit{113}, 969-981. 
\bibitem{roj09} Rojas, O.; Valverde, J.S.; de~Souza, S.M. Generalized transformation for decorated spin models. 
\textit{Physica A} \textbf{2009}, \textit{388}, 1419-1430. 
\bibitem{str10} Stre\v{c}ka, J. Generalized algebraic transformations and exactly solvable classical-quantum models. 
\textit{Phys. Lett. A} \textbf{2010}, \textit{374}, 3718-3722. 
\bibitem{wu71}  Wu, F.Y. Ising Model with Four-Spin Interactions. \textit{Phys. Rev. B} \textbf{1971}, \textit{4}, 2312-2314. 
\bibitem{kad71} Kadanoff, L.P.; Wegner, R.J. Some Critical Properties of the Eight-Vertex Model. \textit{Phys. Rev. B} \textbf{1971}, \textit{4}, 3989-3993.
\end{thebibliography}
\end{document}